\documentclass[useAMS]{mn2e}
\usepackage{psfig}

\usepackage{times}

\def\lsim{ \lower .75ex\hbox{$\sim$} \llap{\raise .27ex \hbox{$<$}} }
\def\gsim{ \lower .75ex \hbox{$\sim$} \llap{\raise .27ex \hbox{$>$}} }

\title[Hard TeV spectrum of 1ES 0229+200] 
{The hard TeV spectrum of 1ES 0229+200: new clues from {\it Swift}}

\author[Tavecchio, Ghisellini, Ghirlanda, Franceschini]
{F. Tavecchio$^1$\thanks{E--mail: fabrizio.tavecchio@brera.inaf.it}, 
G. Ghisellini$^1$, G. Ghirlanda$^1$, L. Costamante$^2$, A. Franceschini $^3$ \\
$^1$ INAF -- Osservatorio Astronomico di Brera, via E. Bianchi 46, I--23807
Merate, Italy\\
$^2$W. W. Hansen Experimental Physics Laboratory and Kavli Institute for Particle Astrophysics and Cosmology, Stanford University, Stanford, CA 94305, USA\\
$^3$ Dipartimento di Astronomia, Universita' di Padova, I--35122, Padova, Italy}

\begin{document}



\maketitle

\begin{abstract} 
The BL Lac object 1ES 0229+200 ($z=0.14$) has been detected by HESS during observations taking place in 2005-2006. The TeV spectrum, when corrected for the absorption of gamma-ray photons through the interaction with the extragalactic background light, is extremely hard, even if the most conservative level for the background is considered. The case of 1ES 0229+200 is very similar to that of 1ES 1101-232, for which a possible explanation, in the framework of the standard one-zone synchrotron-self Compton model, is that the high-energy emission is synchrotron-self Compton radiation of electrons distributed as a power law with a large value of the minimum energy. In this scenario the hard TeV spectrum is accompanied by a very hard synchrotron continuum below the soft X-ray band. We will show that recent {\it Swift} observations of 1ES 0229+200 in the critical UV-X-ray band support this model, showing the presence of the expected spectral break and hard continuum between the UV and the X-ray bands. 
\end{abstract}
 

\section{Introduction}
 
The gamma-ray extragalactic sky at high ($>100$ MeV) and very high ($>100$ GeV) energies is dominated by blazars, associated to radio-loud active galactic nuclei with a relativistic jet closely oriented to the Earth. The resulting relativistic amplification of the non-thermal jet emission makes blazars extreme objects, with apparent luminosities exceeding in the most powerful sources $10^{48}$ erg/s (e.g. Abdo et al. 2009) and variability timescales as short as minutes (e.g. Aharonian et al. 2007a, Albert et al. 2007). The interest for these sources is driven by the possibility to get interesting clues on the acceleration processes of charged particles in relativistic flows and the possible use of their TeV emission to characterize the poorly known extragalactic background light (EBL) in the infrared, optical and UV regions (e.g. Stecker et al. 1992, Stanev \& Franceschini 1998, Costamante et al. 2004, Mazin \& Raue 2007). 

The great majority of blazars detected in the TeV band are BL Lac objects, showing weak or even absent emission lines. Their Spectral Energy Distribution (SED) is characterized by two broad peaks, generally interpreted as due to synchrotron (the low energy peak) and inverse Compton (at high energy) emission (an hadronic origin for the high energy peak is invoked in the hadronic models, e.g. Muecke et al. 2003). Since, as suggested by the weak emission lines, the environment surrounding the jet is thought to be poor of soft photons, the IC emission is likely dominated by the scattering of the synchrotron photons themselves. The synchrotron-self Compton (SSC) scenario is supported by a large number of observational clues. However, some recent observations are difficult to accomodate in this scheme. In particular, one of the most intriguing problems is represented by sources with very hard TeV spectra. Indeed, even considering the lowest level of the EBL allowed by galaxy counts (e.g. Franceschini et al. 2008), some BL Lacs (most notably, 1ES 1101-232 and 1ES 0229+200) have spectra (harder then $F(\nu)\propto \nu^{-1}$, i.e. raising in the $\nu F(\nu)$ representation), challenging the standard SSC scenario, in which the decrease of the scattering cross section with energy inevitably leads to rather soft SSC spectra at TeV energies. 

Different possibilities to produce hard spectra, overcoming the problem of the reduced scattering efficiency, have been  discussed. Katarzy{\'n}ski et al. (2006, K06 hereafter) showed that if the electrons (assumed to follow a power law distribution with energy) are characterized by a relatively large value of the minimum energy, a very hard spectrum can be achieved ($F_{\nu}\propto \nu^{1/3}$).  The same hard spectrum would be visible in the synchrotron component below the soft X-rays.
Stecker et al. (2007) pointed out that, under specific conditions, shock acceleration could produce power law electron distribution much harder than the canonical $\gamma ^{-2}$. We note, however, that even if the electron distribution is hard, the rapid decline of the  efficiency of the inverse Compton scattering at high energy inevitably result in soft spectra. Boettcher et al. (2008) propose to overcome the problem related to the decrease of the scattering cross section assuming that the TeV emission comes from regions of the jet at  distances much larger than those considered in standard models ($<0.1$ pc). At large distances the IC emission should be dominated by the scattering of the extremely soft photons of the CMB (e.g. Tavecchio et al. 2000), for which most of the scatterings happen in the efficient (Thomson) regime. In this model, the X-ray and the TeV emission are decoupled, since the X-rays are assumed to originate in the inner "blazar region". Finally, Aharonian et al. (2008) propose that the absorption of the TeV continuum by a very narrow (black body-like) distribution of optical-UV photons in the source could explain the observed hard TeV spectra. However, there are no indications for the presence of such a very narrow radiation field in the environment of BL Lac objects, though the required field intensities are within the available upper limits.

In this letter we address the problem of the hard TeV spectra for the specific case of 1ES 0229+200 exploiting new data in the UV-X-ray band obtained by {\it Swift}. The UV-X-ray energy range is particularly critical for the model proposed by K06. Indeed, a sharp cut-off is robustly expected in the synchrotron emission at these frequencies, associated to the low-energy cut-off of the electron energy distribution. We use $h_{\rm 0}=\Omega_{\Lambda}\rm =0.7$, $\Omega_{\rm M} = 0.3$.

\section{The SED of 1ES 0229+200}

\subsection{{\it Swift} data}

\begin{figure*}
\vskip -1.5 cm
\hskip -0.5 cm
\psfig{figure=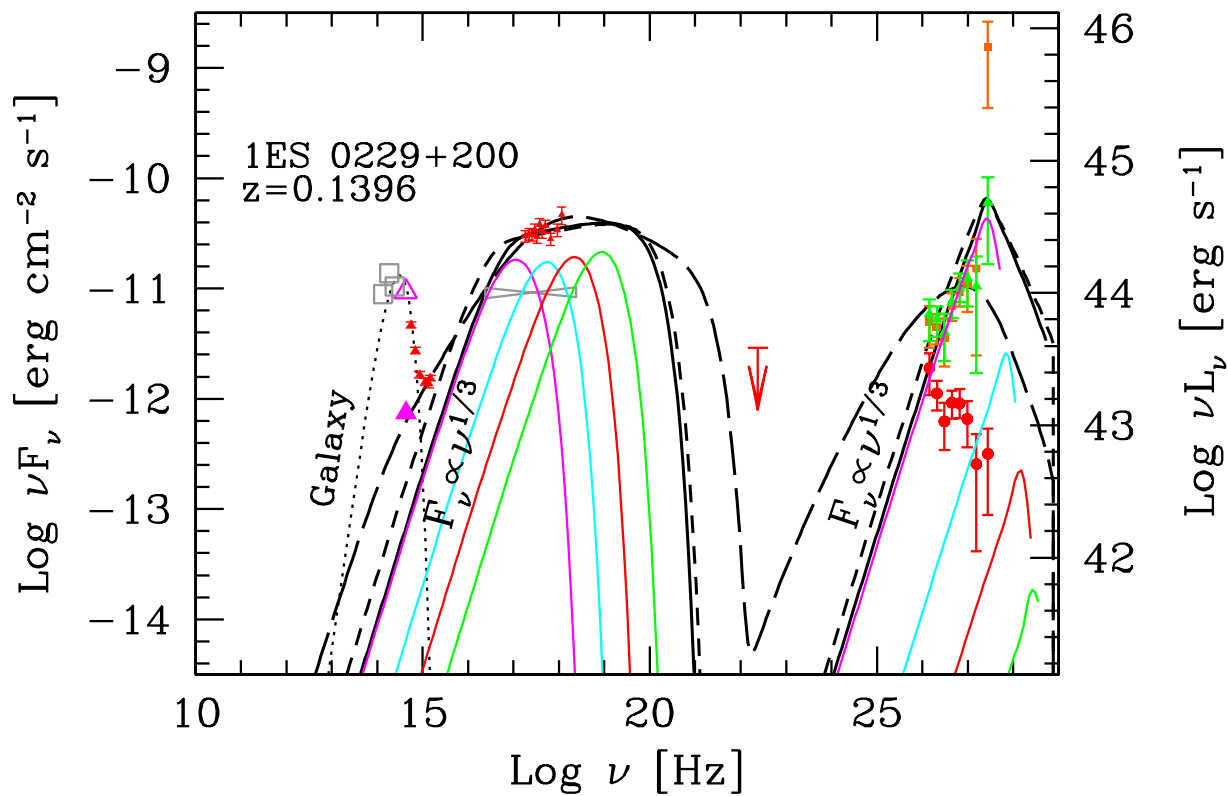,width=17.5cm,height=17.5cm}
\vskip -3 cm
\caption{Spectral energy distribution of 1ES 0229+200. Grey points are from NED. The grey bow-tie is the {\it Beppo}SAX spectrum (Donato et al. 2005). The open magenta triangle (Urry et al. 2000) is the {\it HST} measure for the host galaxy, while the filled magenta triangle is for the core. The red optical-UV and X-ray data are from {\it Swift}. The red arrow shows a rough {\it Fermi} upper limit derived from Abdo et al. (2009). The red filled circles show the observed HESS spectrum (Aharonian et al. 2007b), while the green triangles and the orange squares show the spectrum corrected for absorption by interaction with the EBL calculated with the {\it LowSFR} model of Kneiske (2004) and the model of Franceschini et al. (2008). The black dotted line shows the emission from the host galaxy, approximated with a black body function. The other black lines show the synchrotron and inverse Compton models discussed in the text. For the model shown by the solid black line we also report the 
contribution to the total emission of electrons in small energy ranges (see text for details).
}
\label{sed}
\end{figure*}

{\it Swift} (Gehrels et al. 2004) observed 1ES 0229+200 two times, on 2008  August 7 and 8, with a net exposure of 3.2 and 3.3 ksec, respectively.  XRT (Burrows et al. 2005) and UVOT (Roming et al. 2005) data have been reduced using the standard procedure (described in, e.g., Ghisellini et al. 2007), with the latest version of the \texttt{HEASOFT} and \texttt{CALDB} packages\footnote{http://heasarc.gsfc.nasa.gov/docs/software/lheasoft/}. XRT data have been fitted with \texttt{XSPEC12}.
The data of the first observation are better fitted with a broken power law model with a very hard low energy slope (Table \ref{xrt}), while the second data set is well fitted with a single power law. However, above 1 keV the slope and the flux level are very similar. For simplicity, in the SED (Fig. \ref{sed}) we only report the data corresponding to the second observation.
UVOT magnitudes have been corrected for Galactic absorption using the values of Schlegel et al. (1998) for the V, B and U filters and the formulae by Pei (1992) for the UV filters and converted into fluxes following Poole et al. (2008).

\begin{table}
\begin{tabular}{lccccc} 
\hline  
Date   & $\Gamma _1$&  $\Gamma _2$   &  $E_b$  & F$^a$ & $\chi^2/$d.o.f. \\
\hline 
7/8/2008 &$0.7^{+0.7}_{-1.3}$ &$1.93\pm 0.15$ & $0.85\pm0.25$& 1.00& 23/19\\
8/8/2008& $1.8\pm 0.1$& -- & --&1.03 &23.6/21\\
\hline \\
\end{tabular}
\caption{Fit of the XRT data of 7/8/2008 (8/8/2008) with the broken power law (power law) model. $^a$: deabsorbed flux in the 0.3-10 keV band in units of $10^{-10}$ erg cm$^{-2}$ s$^{-1}$. In both cases the hydrogen column density is fixed to the Galactic value, $N_H=9.2\times 10^{20}$ cm$^{-2}$.}
\label{xrt}
\end{table}

\subsection{TeV data}

The HESS spectrum has been measured in 2005-2006 by Aharonian et al. (2007b, red circles in Fig. \ref{sed}). For such distant blazar detected at such high energies, the correction of the observed TeV data for the photon-photon opacity is most relevant.

In the case of 1ES 0229+200, its redshift implies that the energy photons detected at 10 TeV mainly  interact with cosmic background photons with wavelength $\lambda_{max} \simeq 1.24 (E_\gamma [TeV])\ \mu$m.
The maximum observed energy of TeV photons is $\sim$11 TeV, corresponding to 12.5 TeV in the source rest frame. 
So the bulk of the absorption is produced by background photons around 13-15 $\mu$m. At these wavelengths the IR background is not directly measurable because of the exceedingly large intensity of the inter-planetary dust foreground. However the properties of the cosmic sources of these photons are known in great detail from three independent survey missions: IRAS (observing at 12 and 25 $\mu$m), ISO with maximal sensitivity at 15 $\mu$m, and Spitzer at 8 and 24 $\mu$m. Also various independent approaches can be used to estimate the background intensities. The simplest and most robust one is to determine the source number counts over the widest flux density range at various wavelengths and to integrate them to get the local background intensity. The other solution is to calculate the multi-wavelength luminosity functions at different cosmic epochs (redshift intervals), a quite more complex procedure, but suited for computing the redshift evolution of the background photon density. Note that all these procedures yield lower limits 
 to the background intensity.

We have then considered two background levels for de-absorption correction. The first one uses the "Low-SFR" model of Kneiske et al. (2004), providing minimal IR background level and absorption correction. TeV data corrected in such way are reported in Fig. 1 as green triangles.
We have also computed a corrected spectrum (orange squares) using the observationally-based model of Franceschini et al. (2008), which takes into account all available data on the background sources. The model includes also some corrections related to the evolution of the background intensity as mentioned above (the latter are small, however, for our source's redshift, and tend in any case to reduce the absorption correction).

The corresponding intrinsic spectrum, even with the conservative EBL level assumed in the models, is very hard  (photon index $\sim 1.5\pm 0.2$) and the high energy peak in the SED appears to be located above 10 TeV. Particularly critical appears the last data point (at energy $E\simeq 11$ TeV): once corrected with the Franceschini et al. (2008) model, providing a much larger optical depth at these energies than the Kneiske et al. (2004) model, the point lies at a flux that it is very difficult to reconcile with the current emission models 
(in a similar way to Mkn 501 around 15-20 TeV, see e.g. Aharonian 2001). However, the very low statistics at the highest 
energy bins and the uncertainty  on the EBL spectrum at 10-20 microns do not allow  yet any reliable conclusion. 
Some caveats also apply to the spectral shape at smaller energies. Indeed, although the spectrum is well represented by a power law, at a closer look it hints at a more complex shape (this is already visible in the deabsorbed spectrum shown in Aharonian et al. 2007b). In particular, the first bins seems to follow a relatively soft power law, while the datapoints above 2 TeV describe a harder spectrum (or even a bump). This feature in the average spectrum could be either due to variability in the spectral shape or  could indicate that more than one component contributes to the total emission. However, it also depends on the specific shape of the EBL spectrum between 1 and 10 micron. For simplicity, and because this feature is  not significant, in the following  we assume that the intrinsic source spectrum  is represented by a single hard power law.

\subsection{The SED}

We assembled the SED of 1ES 0229+200 (Fig. \ref{sed}) using historical optical-IR data (from NED and Urry et al. 2000),  the simultaneous (de-reddened) optical-UV and X-ray data from {\it Swift} and the de-absorbed HESS data points (average of measures taken in 2005-2006). The red arrow shows the {\it Fermi} upper limit inferred from the absence of 1ES 0229+200 from the list of high-confidence sources detected during August-October 2008, the period of the first three months of observations, reported in Abdo et al. (2009).

As shown in Urry et al. (2000) using {\it HST} data, the optical flux is dominated by the host galaxy (open magenta triangle), while the flux of the core lies almost an order of magnitude below that of the host (filled triangle). Therefore, it is very likely that also the IR-optical data from the 2 MASS  survey (open gray squares)  are dominated by the host galaxy. This possibility seems to be confirmed by the UVOT data (red triangles), showing a rapid decrease of the emission from the optical to the UV band. We can therefore conclude that the optical (and partly also the UV) datapoints are dominated by the emission from the galaxy (approximated as a black body in Fig. \ref{sed}, dotted line) and that the optical-UV non-thermal continuum of the jet lies below the UVOT datapoints. 

Unfortunately the HESS data are not simultaneous to the {\it Swift} pointing. However, as discussed by Aharonian et al. (2007b), the TeV spectrum of 1ES 0229+200 showed a quite limited variability during the observations in 2005-2006 (small variability also characterizes the other TeV blazar with a very hard spectrum, 1ES 1101-232). Therefore we think that we can reasonably assume that the TeV spectrum is similar to the 2005-2006 average HESS spectrum.

\section{Modeling the SED}

We reproduce the overall SED with the one-zone synchrotron SSC model fully described in Maraschi \& Tavecchio (2003). Briefly, the emitting region is a sphere with radius $R$ moving with bulk Lorentz factor $\Gamma$, with a tangled and uniform magnetic field $B$. The viewing angle of the observer is $\theta$, which, in turn, means a Doppler factor $\delta$. In the original model the electrons are assumed to follow a smooth broken power law energy distribution
with normalization $K$ and indices $n$ from $\gamma_{\rm min}$ to $\gamma_{\rm break}$ and $n_2$ above the break up to $\gamma_{\rm max}$. For some of the applications in this paper (specifically, models 2 and 3 below) we also use a single power law, with slope $n$ and limits $\gamma_{\rm min}$  and $\gamma_{\rm max}$ and normalization $K$ (extrapolated at $\gamma =1$). 

A ''standard" SSC model with a broken power law electron energy distribution extending down to relatively low energies (model 1 in Table \ref{model}) fails in reproducing the hard TeV spectrum. The main difficulty is to locate the peak of the SSC bump at energy as large as few TeV, as required by the data. As detailed in Tavecchio \& Ghisellini (2008), if one requires a large separation between the synchrotron and the SSC peaks a very large Doppler factor is necessary. Even with $\delta=50$ (see model 1 in Table \ref{model}), the SSC peak lies well below 10 TeV. Note also that, in this case, in order to not overproduce the observed UV-optical flux we are forced to use a rather hard electron distribution, $n=1.5$, harder than the canonical value of 2 (possibilities to have such hard spectra have been discussed, see e.g. Stecker et al. 2007 and Virtanen \& Vainio 2005).

\begin{table*}
\begin{center}  
\begin{tabular}{lccccccccc} 
\hline  
     &$\gamma _{\rm min}$ &$ \gamma_{\rm b}$ & $\gamma _{\rm max}$& $n$&  $n_2$ & $B$ [G]& $K$ [cm$^{-3}$]& $R$ [cm]& $\delta $ \\
\hline 
1    & $10^4$& $6\times 10^5$& $3\times 10^7$& 1.5& 3.4& $8.5\times 10^{-3}$& 6& $10^{16}$&50  \\
2    & $8.5\times 10^5$ & -- &$4\times 10^7$& 2.85 &-- & $5\times 10^{-4}$& $3.5\times 10^9$& $5.4\times 10^{16}$&30  \\
3     & $5\times 10^5$ & -- &$4\times 10^7$& 2.85 &-- & $4\times 10^{-4}$& $6.7\times 10^8$& $5.4\times 10^{16}$&50  \\
\hline \\
\end{tabular}
\end{center}
\caption{Input parameters for the models shown in Fig.\ref{sed} (model 1: long dashed line; model 2: solid line; model 3: short dashed line). See text for definitions.}
\label{model}
\end{table*}

The shape of the SED in the optical-X-ray region shows that the synchrotron continuum falls quite rapidly from the X-ray to the UV band. As already mentioned this is what  we would expect in the case of a truncated electron energy distribution, with the emission being the low-energy tail of the electrons at the lowest energy. It is thus tempting to model the entire SED with such a model. The solid line in Fig. \ref{sed} shows the model assuming a single power law with a large minimum electron Lorentz factor (model 2 in Table \ref{model}), $\gamma _{\rm min}=8.5\times 10^5$. As discussed in K06, below the typical frequency of the electrons at $\gamma _{\rm min}$  the synchrotron spectrum is very hard, showing the characteristics slope $F_{\nu }\propto \nu ^{1/3}$.  The corresponding SSC spectrum shows a peculiar "triangular" shape, with a low energy power law reflecting the  $F_{\nu }\propto \nu ^{1/3}$ slope of the synchrotron target photons, and a fast decrease after the maximum due to the rapid decline of the scattering efficiency. 

To better visualize the situation, we also report (colored curves) the synchrotron and the SSC emission of electrons in small "slices" of energy, with Lorentz factors in the range $8.5-20\times 10^5$ (magenta), $2-4\times 10^6$ (cyan), $4-8\times 10^6$ (red) and $8-16\times 10^6$ (green). As can be seen, the SSC emission is almost entirely dominated by the emission from the low energy end of the distribution (magenta), also responsible for the hard tail of the synchrotron component. While the luminosity of the synchrotron emission from electrons of different energies is almost equal, 
the decrease of the scattering efficiency makes the luminosity of the SSC emission to strongly decrease for electrons with increasing energy.

The position of the SSC peak in the Klein-Nishina regime reflects the energy of the electrons at the low energy end, $\nu _{\rm IC}=\gamma _{\rm min} (mc^2/h) \delta $ (e.g. Tavecchio et al. 1998). The corresponding synchrotron frequency is $\nu _{\rm s} \simeq 3.7\times 10^6 B \gamma _{\rm min}^2 \delta$. Then the ratio between the magnetic field and the Doppler factor:
\begin{equation}
\frac{B}{\delta} = 1.7 \times 10^{-2} \frac {E _{\rm s, keV}}{E_{\rm IC, TeV}^2}  \,\,\,\ {\rm G}
\label{bsudelta}
\end{equation}
\noindent
where $E _{\rm s, keV}=h\nu _{\rm s}$ is measured in keV and $E_{\rm IC, TeV}=h\nu _{\rm IC}$, the energy of the SSC peak, is in  TeV. Setting $E _{\rm s}\simeq 0.1$, $E_{\rm IC}\simeq 10$ and $\delta=30$, we find $B=0.5$ mG, much smaller than what is usually found in other TeV BL Lacs (i.e. $B\sim 0.1-1$; e.g. Celotti \& Ghisellini 2008). As detailed in Tavecchio \& Ghisellini (2008), to increase the separation between the two peaks (i.e. move the SSC peak at larger energies and/or the synchrotron peak at lower energies) we have to increase the Doppler factor. As an example of this, we report with the short dashed line another possible model with the same SSC peak frequency and a lower synchrotron break frequency. In this case the required value of Doppler factor is $\delta=50 $.

\section{Discussion}

The combination of the optical-UV  and X-ray data of 1ES 0229+200 suggests a rapid increase of the flux between the UV and the X-ray band, that we interpret (following K06) as due to a large value of the high-energy cut-off in the electron energy distribution, leading to a very hard synchrotron continuum, $F_{\nu }\propto \nu ^{1/3}$. This synchrotron spectrum translates into a similarly hard SSC spectrum, reproducing the intrinsic (de-absorbed) TeV data. 
We note that, at a closer look, the spectrum predicted by the model seems to be harder than what required by the data derived using both the considered models for the EBL. A better agreement would be attained by considering larger levels for the EBL, such as that adopted in the "Best model" of Kneiske et al. (2004). However, the detection of blazars at large redshift at TeV energies (Aharonian et al. 2006, Albert et al. 2008) seems to exclude such high levels for the EBL.

We remark that, within the framework of the one-zone SSC model, the solution discussed here is the only one able to reproduce hard TeV spectra, without invoking unrealistically large Doppler factors. As we have shown, this solution has also a direct consequence on the synchrotron continuum, predicting a sharp roll-off of the emission below the X-ray band. The UV-X-ray data from {\it Swift}, though not simultaneous with the TeV spectrum of HESS, strongly support our scenario.

The model has some important consequences for the acceleration process and the role of the energy losses on the electrons. This is related to the requirement of an of the electron energy distribution with a sharp low-energy end. If there were electrons below this limit, even with a very hard distribution, the resulting SSC spectrum would be completely different: the peak, dominated by  the emission of the low-energy electrons (see Tavecchio et al. 1998), would shift at energies well below 10 TeV (see also K06). Therefore it is essential to assume that (i) the acceleration process is able to push almost all the electrons at energies above $\gamma _{\rm min}$ and (ii) that the electrons, once  injected into the emission region, do not cool rapidly. In fact, the cooling would inevitably lead to the formation of the standard $\gamma ^{-2}$ power law below $\gamma _{\rm min}$.  The requirement (i) is particularly demanding for the models based on relativistic shocks. This could indicate that electrons are powered by reconnection of magnetic fields (e.g. Giannios, Uzdenky \& Begelman 2009 for an application to blazars). This conclusion is supported by the extremely low value of the magnetic field derived in the emission region. Assumption (ii) is instead naturally accomplished in our model: with the parameters used for model 2 we derive a (observed) cooling time of the electrons at $\gamma _{min}$  of $t_{\rm cool}\simeq 5\times 10^8/\gamma _{\rm min}B^2 \delta \simeq 10^8\; {\rm s} \sim 47 \; {\rm months}$. Therefore, such a configuration could be maintained for some years. This cooling timescale is valid if only radiative cooling are efficient. Realistically, the emitting source is expanding, thus adiabatic losses are more efficient in cooling the electrons. The stability of the TeV flux for long timescales suggest than that the acceleration/injection mechanism can act for long times at a constant level.

A direct consequence of this scenario is the very low level of the emission in the GeV band. Thus a prediction is that {\it Fermi} will not detect 1ES 0229+200. It is possible (or even likely) that there are other active regions in the jet, possibly producing $\gamma $-rays in the {\it Fermi} range. In this case, since the emission in the TeV and in the GeV band would not come from the same location, a strict correlation between the variations in the two bands is not expected. 

Our model for the emission of 1ES 0229+200 reproduces both the hard UV-soft X-ray continuum and the very hard TeV spectrum. In the already mentioned model of Boettcher et al. (2008), instead, the UV-X-ray and the TeV components are not related, being emitted in very different regions of the jet (X-rays in the blazar region, TeV emission in the large scale jet). Therefore a very effective way to distinguish between these two alternatives is through variability of the optical-X-ray and TeV emissions. Related variations would point to a common region for the two emissions, challenging the model of Boettcher et al. (2008). On the contrary, totally independent variability would indicate that two regions are active, ruling out our scenario (and any one-zone model). Moreover, Boettcher et al. (2008) predict a larger GeV flux than predicted in our scenario, coming from the SSC emission of the inner region responsible for the X-ray continuum (albeit the GeV emission could be accommodated with a different choice of parameters). 
Another possible criticism could be that in order to fit the data, $\delta=25$ is required in the external regions of the jet, contrary to the VLBI indications, suggesting very small apparent velocity at parsec scales (e.g. Piner, Pants \& Edwards 2008). The same problem could afflict our model: a strong deceleration from $\Gamma =15-30$ in the emission region to $\Gamma\sim $a few at parsec scale would imply a radiative luminosity comparable to the jet power (see also Ghisellini, Tavecchio \& Chiaberge 2005). On the contrary, the derived jet power is more than one order of magnitude larger than the radiative power. Unfortunately there are no measurements of the jet speed in 1ES 0229+200 at VLBI scale.

Concluding, the model presented here is capable of reproducing successfully the main features of the observed SED, even if implying extreme physical parameters, in particular for the minimum electron energy and the magnetic field. These conditions are rather different than those commonly derived in other TeV BL Lacs. This could indicate physical conditions in the jet more complex than what assumed in the simple one-zone model adopted here, requiring more sophisticated models (multi-zone, multi-population of particles).

A rather interesting point concerns the consequences of the observed TeV spectrum for the modeling of the EBL. Even using the strict lower limits coming from the galaxy counts the derived intrinsic spectrum is exceptionally hard, with the point at the largest energy being the most problematic. Further observations of such hard spectra, in this and other sources, particularly if extended at large energies ($>10$ TeV), could then be particularly relevant, possibly leading to the exciting evidence that something fundamental is missing in our understanding of the emission physics of blazars or of the propagation of $\gamma$-ray photons in the Universe (e.g. De Angelis et al. 2009).

\section*{Acknowledgments}
We thank the referee for the helpful report.
This work was partly financially supported by a 2007 COFIN-MIUR grant. This research has made use of the NASA/IPAC Extragalactic Database (NED) which is operated by the Jet Propulsion Laboratory, Caltec, under contract with the NASA. We acknowledge the use of public data from the Swift data archive. This research has made use of data obtained from the High Energy Astrophysics Science Archive Research Center (HEASARC), provided by NASAÕs GSFC. Support from ASI under contract ASI/DA016, is acknowledged.

\end{document}